# On the Challenges and KPIs for Benchmarking Open-Source NFV MANO Systems: OSM vs ONAP

Girma M. Yilma, Faqir Zarrar Yousaf, Vincenzo Sciancalepore, Xavier Costa-Perez
NEC Laboratories Europe GmbH, Germany
Email: {girma.yilma|zarrar.yousaf|vincenzo.sciancalepore|xavier.costa}@neclab.eu

*Abstract*—NFV management and orchestration (MANO) systems are being developed to meet the agile and flexible management requirements of virtualized network services in the 5G era and beyond. In this regard, ETSI ISG NFV has specified a standard NFV MANO system that is being used as a reference by MANO system vendors as well as open-source MANO projects. However, in the absence of MANO specific KPIs, it is difficult for users to make an informed decision on the choice of the MANO system better suited to meet their needs. Given the absence of any formal MANO specific KPIs on the basis of which a performance of a MANO system can be quantified, benchmarked and compared, users are left with simply comparing the claimed feature set. It is thus the motivation of this paper to highlight the challenges of testing and validating MANO systems in general, and propose MANO specific KPIs. Based on the proposed KPIs, we analyze and compare the performance of the two most popular open-source MANO projects, namely ONAP and OSM, using a complex open-source vCPE VNF and identify the features/performance gaps. In addition, we also provide a sketch of a test-jig that has been designed for benchmarking MANO systems.

## I. INTRODUCTION

AGILITY and flexibility for the management of the network resources and services represents one of the key innovations of 5G networks to support carrier-grade operations for different verticals with diverse service requirements at reduced CAPEX/OPEX costs. In this context, Network Function Virtualization (NFV) has been widely accepted as a technology enabler for addressing the challenging requirements of 5G networks [1]. The key concept of NFV is the decoupling of the network functions from the underlying hardware platforms, while the network functions are realized as a virtualized entity commonly referred to as Virtualized Network Functions (VNFs). VNFs can embody less complex network functions such as Firewall (vFW), load balancer (vLB) to more complex functions such as Evolved Packet Core (vEPC), Customer Premises Equipment (vCPE) to name a few. End-to-end Network Services (NS) are composed by chaining relevant VNFs over Virtual Links (VL).

The introduction of NFV technology has great implications on the network management systems where they need to be extended to provide Life Cycle Management (LCM) of VNFs, NSs and VLs beyond the traditional FCAPS (Fault, Configuration, Accounting, Performance, Security) management services. The LCM actions include operations such as on-boarding, instantiation, scaling in/out/up/down, migration, update/upgrade, etc of a VNF and its associated components.

In this regard the ETSI ISG NFV has proposed a standard NFV Management and Orchestration (MANO) framework [2] and has specified interfaces and operations on its various reference points to support different functional features in its various specification documents. Fig. 1 provides a high level overview of the ETSI NFV MANO system functional blocks and the various interfaces defined on the reference points. The ETSI NFV MANO framework is also serving as a reference to other independent MANO projects that are being undertaken either by vendors or by open source communities. The latter is gaining a lot of prominence and attention from operators due to the diverse efforts that are being expended towards developing open source MANO platforms.

### A. Problem Statement

Open source MANO projects such as ONAP [3], OSM [4], Open Baton [5], Cloudify [6], OPNFV [7], are under different stages of steady development. All are competing to make their mark in the operators' infrastructure but, owing to the complex nature of the NFV MANO system itself, no project to date can claim to support the entire LCM spectrum of the NFV assets or be ready for field operations. More prominent among these projects are Open Network Automation Platform (ONAP) and Open Source MANO (OSM), which have gained a lot of attention from the operators' community, especially because of the patronage of some big operators behind the development of ONAP and OSM. For instance ONAP, which is being developed under the umbrella of the Linux Foundation, is mainly supported by AT&T, whereas OSM is driven by Telefonica and is being developed under the mandate of the newly formed ETSI Open Source Group (OSG).

Both ONAP and OSM are under different stages of their releases but they are far from being complete or stable. Both are aiming to provide an integrated NFV MANO framework, but they are following very different directions in terms of architecture and implementation. *There are still gaps between what is being claimed and what features and functionalities are actually supported*. There are ambiguities in terms of their deployment footprint as well as operational efficiency for providing carrier-grade management to NFV services. Owing to the fact that these are relatively latest developments, there is very much less information and experience available in terms of the functional and operational capabilities of these platforms and technology readiness level (TRL).

Moreover, carrying out performance benchmarking of MANO systems is in itself a challenge. This is because unlike other traditional network entities, that have well defined KPIs to benchmark the performance, there are no set and well defined KPIs on the basis of which the performance of a MANO system can be benchmarked. In view of this, the main objective of this paper is to (1) highlight the KPIs on the basis of which a MANO system performance can be analyzed, (2) to analyze and compare the performance of ONAP and OSM MANO frameworks based on the identified KPIs, and (3) provide a sketch of a test-jig that has been established for benchmarking MANO systems.

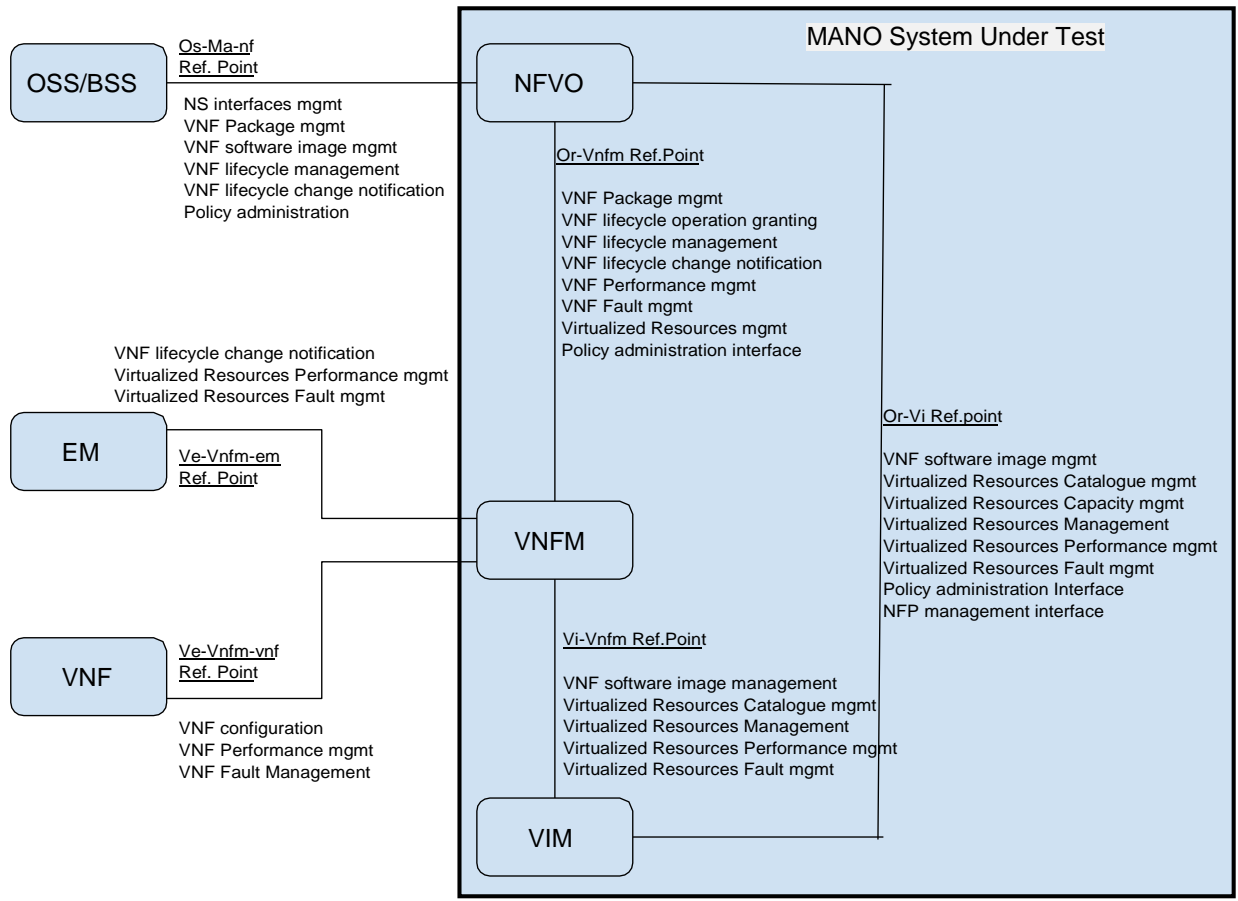

Figure 1: Interface Mapping to ETSI NFV MANO Reference Points.

The rest of the paper is organized as follows. The next Section II will provide the related work while in Section III we highlight the challenges in analyzing the performance of a MANO system and provide the base KPIs and methodology for such an analysis. Based on the KPIs, we will then proceed to compare the performance of ONAP and OSM and provide the analysis in Section IV. Since ONAP and OSM are in a continuous state of development, we will provide concluding remarks in Section V, and finally we point directions for the future works in Section VI

## II. RELATED WORK

An NFV ecosystem offers a complex mix of management and managed entities. The NFV MANO system entities, such as the Network Function Virtualization Orchestrator (NFVO), the Virtual Network Function Manager (VNFM) and the Virtual Infrastructure Manager (VIM), coordinate with each other over well-defined reference points to manage entities such as Network Functions Virtualization Infrastructure (NFVI), VNFs and NSs. In the space of performance testing and analysis of managed entities (i.e., NFVI components, VNFs and NSs) there are clear guidelines and experience available with well-defined KPIs and test processes both in the literature and standard's body such as ETSI ISG NFV.

In the *research space* the performance analysis is limited to managed assets such as NFVIs, VNFs and NSs. For example, in [8] the authors highlight the differences in testing of VNFs from Physical Network Functions (PNFs), where the latter are bundled on dedicated hardware platforms. In this context the authors introduced the concept of "white-box testing", where measurements are made for each component in the ecosystem, which is in contrast to the "black-box testing" which is the common practice for benchmarking network devices. Another notable work in [9] compares the performance of three different VIM platforms, namely OpenStack [10], OpenVIM [11] and Nomad [12], in terms of the instantiation process of Virtual Machines(VMs). Another good work in [13] motivates the need for community-driven network service benchmarking. They extend the OPNFV Yardstick framework to perform real-world VNF and NFVI characterization and benchmarking with repeatable and deterministic methods. In [14] a platform for VNF benchmarking as a service is proposed. The authors claim their approach enables not only run-time resource eval-uation but also test-before-deploy opportunities for VNFs and NFVIs.

In *standards' domain*, the Testing Working Group (TST WG) in ETSI ISG NFV is working on specifying performance metrics, and providing guidelines on test methodologies to benchmark the performance of the different aspects of the NFV MANO ecosystem.

The ETSI GS NFV-TST 007 specification [15] describes a set of informative interoperability test guidelines for NFV capabilities that require interactions between the components implementing NFV functionality, namely, the NFVO, VNFM, EM-VNF and VIM-NFVI. It provides detailed guidelines on *functional testing* of the various supported features and interfaces. The result is a binary indicating whether a particular feature or interface operation on a specified reference point is supported or not.

The ETSI GS NFV-TST 008 specification [16] is NFVI specific and specifies detailed and vendor-agnostic key operational performance metrics at different layers of the NFVI, especially processor usage and network interface usage metrics. The identified performance metrics are associated with the compute nodes of the NFVI and are well-known such as processor utilization, packet count, dropped packet count, memory utilization etc.

The ETSI GS NFV-TST 009 specification [17] is also NFVI-specific and specifies vendor-agnostic definitions of performance metrics and the associated methods of measurement for benchmarking networks supported in the NFVI for fair comparison of different implementations of NFVI. The KPIs are typically network related such as throughput, latency, delay variation and loss.

As is evident, the activities in both the research domain and ETSI NFV is based on specifying KPIs and testing methodologies for NFVI related resources, and from the NFV MANO perspective only specifies feature testing.

Recently under the EU's H2020 SONATA project a deliverable was published [18] that validated the performance of the developed SONATA MANO SDK. For validation they used metrics such as NS implementation and creation effort, NS packaging time, the test environment setup time, service deployment time and service instantiation time. However, they did not specify or take into account metrics for run-time LCM operations. Moreover, except the service deployment and instantiation time, the other metrics pertain to a one-time operation of packaging and on-boarding the software images, and thus are less important when considering the day-to-day active MANO system operations.

In [19] and [20], a MANO specific KPI is defined which is referred to as *Quality of Decision (QoD)*. QoD is a measure of a MANO's system capabilities in terms of *deriving* optimized LCM decisions and has been used as a reference KPI in [21] for analyzing the performance of a z-TOrch orchestration algorithm for optimizing the LCM decisions of NFV MANO system with reduced monitoring load. However, it still does not measure the performance of the respective LCM operations.

In view of the existing work and the gaps pointed above, we specify and define both the *Functional* and *Operational* KPIs that can be used for analyzing and benchmarking the management performance of NFV MANO platforms in the next section.

III. NFV MANO SYSTEMS BENCHMARKING: CHALLENGES & KPIS

In this section we highlight the need and challenges to benchmark the performance of NFV MANO systems. We then propose KPIs that can be used as a basis for analyzing the performance of NFV MANO systems.

*A. MANO Benchmarking Challenges*

Traditional Network Management System (NMS) is based on FCAPS management that involves the NMS to monitor, measure and enforce the KPIs of the networks and services that they are managing. Some of the key KPIs that an NMS monitors are:

1) Availability – to ensure that the network elements and services are available.
2) Utilization – to ensure that network resources are utilized to their maximum
3) Service Level Agreements (SLA) – to ensure that users and services do not exceed their respective utilization of resources beyond what is stipulated.
4) Latency – to ensure that the services are delivered within the specified delay budget to maintain Quality of Service (QoS) and Quality of Experience (QoE).
5) Jitter – to ensure that the packet inter-arrival time does not deviate from the mean delay beyond a specified degree.
6) Errors – to ensure the maintenance of end-to-end service integrity against errors.

There are many other measurable parameters that fall within the above mentioned categories of KPIs that an NMS monitors and enforces by taking appropriate measures. The above mentioned KPIs are well-known, well defined and they also cover the performance measurements of the NFVI compute and network resources as well as the VNFs. For instance, the performance of a virtualized router function can be benchmarked using the traditional KPIs defined for the traditional routers.

The challenge however is to define the KPIs for benchmarking the performance of the NFV MANO system itself. This becomes all the more important because the NFV MANO system, in addition to performing the traditional FCAPS management, provides LCM of VNFs and NSs. Having KPIs to quantify the performance of MANO systems is important owing to the highly agile, flexible and dynamic nature of the virtualized services being delivered by the VNFs/NSs where the reaction time of the NFV MANO system from monitoring an event to deriving an appropriate LCM action and executing it becomes critical in view of stringent performance requirements of different verticals sharing the same NFVI resources. A benchmarked MANO system will thus provide the customer to choose an appropriate MANO solution that best fits its operational needs.

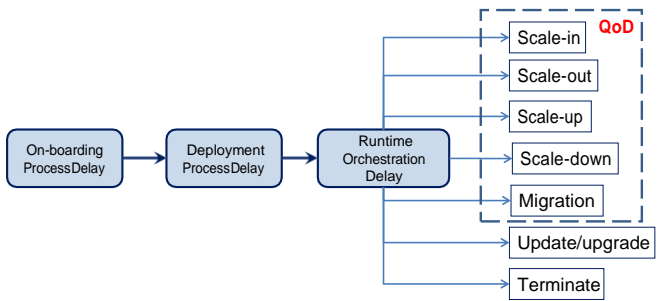

Figure 2: KPIs for MANO Life Cycle Management Operations.

*B. MANO Performance KPIs*

We propose to classify the MANO system performance KPIs into two categories, namely the *Functional KPIs* and *Operational KPIs*.

*Functional KPIs* describe non-run-time characteristics of a MANO system. These include:

- Resource footprint of the deployed MANO system.
- Variety of VIM platforms a MANO system can support.
- Number of VIMs a single MANO platform can manage efficiently.
- The maximum number of VNFs a MANO system can effectively/efficiently monitor and manage in an NFVI.
- Feature palette. For example, support for DevOps, VNF Package Management, VNF image management, integrated monitoring system, etc.

*Operational KPIs* characterize the run-time operations. This is mainly quantified by measuring the time-latency of a particular LCM action/task and its effectiveness. These are depicted in Fig. 2 and summarized below:

1) **On-boarding Process Delay (OPD):** This is the time that it takes to boot-up a virtualized network function image i.e., a VM with all its resources. Once booted, the VM will be used to host and run a VNF service. This is similar to the service deployment time defined in [18].
   The prerequisite before on-boarding a VM is the creation of a VNF software image in a format that is recognized by the MANO system, and the package uploaded to a repository. This package not only contains the VNF software but also the VNF descriptor file (VNFD) that specifies all the configuration information, network requirements, resource requirement, routing/security policy, IP ranges, performance requirement, interfaces etc. A NS Descriptor (NSD) is also on-boarded which, in simpler terms, is a template describing the NS in terms of its functional, operational, performance, security, links, QoS, QoE, reliability, connectivity, requirements. It also includes the VNF Forwarding Graph (VNFFG) that identifies the VNF types and the order of their connectivity and the characteristics of the VLs interconnecting the constituent VNFs to create a NS. OPD is dependent on the service resource requirements specified inside the VNFD.

2) **Deployment Process Delay (DPD):** This is the time it takes to deploy and instantiate a VNF within the booted VM and setup an operational NS. In this process, a service instance is instantiated by parsing the NSD and the VNFD files. All the VNFs that are part of the NS are instantiated based on their respective on-boarded images. The MANO system will ensure the provision of required resources for instantiating the VNFs and linking them via relevant VLs in case of a complex NS, and then configure each VNF based on the configuration information provided in the respective NSDs and VNFDs. The speed at which a VNF or a NS is deployed is crucial when a VNF/NS has to be scaled to meet sudden increase in load demands. This is somewhat similar to the network service instantiation time mentioned in [18].

3) **Run-time Orchestration Delay (ROD):** As shown in Fig. 2 the run-time orchestration operations consist of different management actions, and the performance latency of each individual action can be quantified by measuring the time difference from the moment the action is executed to the time the action is completed. For example, a MANO system that can complete a scale-out or a migration operation of a heavily loaded VNF with minimum service disruption can be deemed to have good performance. ROD is dependent on a monitoring system that continuously monitors active VNF/NS instances throughout their lifetime for any performance deviation or fault event. Thus a MANO system that performs low-latency run-time operations with minimum monitoring load can be considered as performant.

4) **Quality-of-Decision (QoD):** QoD is another metric that was introduced in [19] and demonstrated in [20] and [21], where it quantifies the performance of a MANO system in terms of its effectiveness in carrying out run-time LCM operations of VNF scaling and migration (see Fig. 2). QoD is measured in terms of the following criteria:
   a) Efficiency of a resource management decision. The resource efficiency is measured in terms of:
      i) Whether both the long-term and short-term resource requirements of the managed VNF will be fulfilled in the selected compute node.
      ii) How non-intrusive a management action has been for other VNFs that are already provisioned in the selected compute node. That is, to what extent will the managed VNF VM affect the performance of other VNFs in the selected compute node in terms of resource availability.
   b) Number of times the management action has to be executed before the most-suitable compute node is determined to migrate/scale the managed VNF to.
   c) The timeliness of the computation and execution of MANO LCM actions. The latter criterion being relevant more pronounced in the management of a multi-site NS scenario as described below.

IV. OPEN-SOURCE NFV MANO SYSTEMS PERFORMANCE EVALUATION: OSM VS ONAP

In this section we will analyze and compare the performance of two popular but competing open source MANO platforms namely ONAP and OSM. We will analyze the performance of these two platforms based on the KPIs proposed in the previous section in order to ascertain not only their performance but also their technology readiness level (TRL) in view of active deployment in production networks. We will first describe the test-bed set up and the test methodology, and afterwards present results and their analysis.

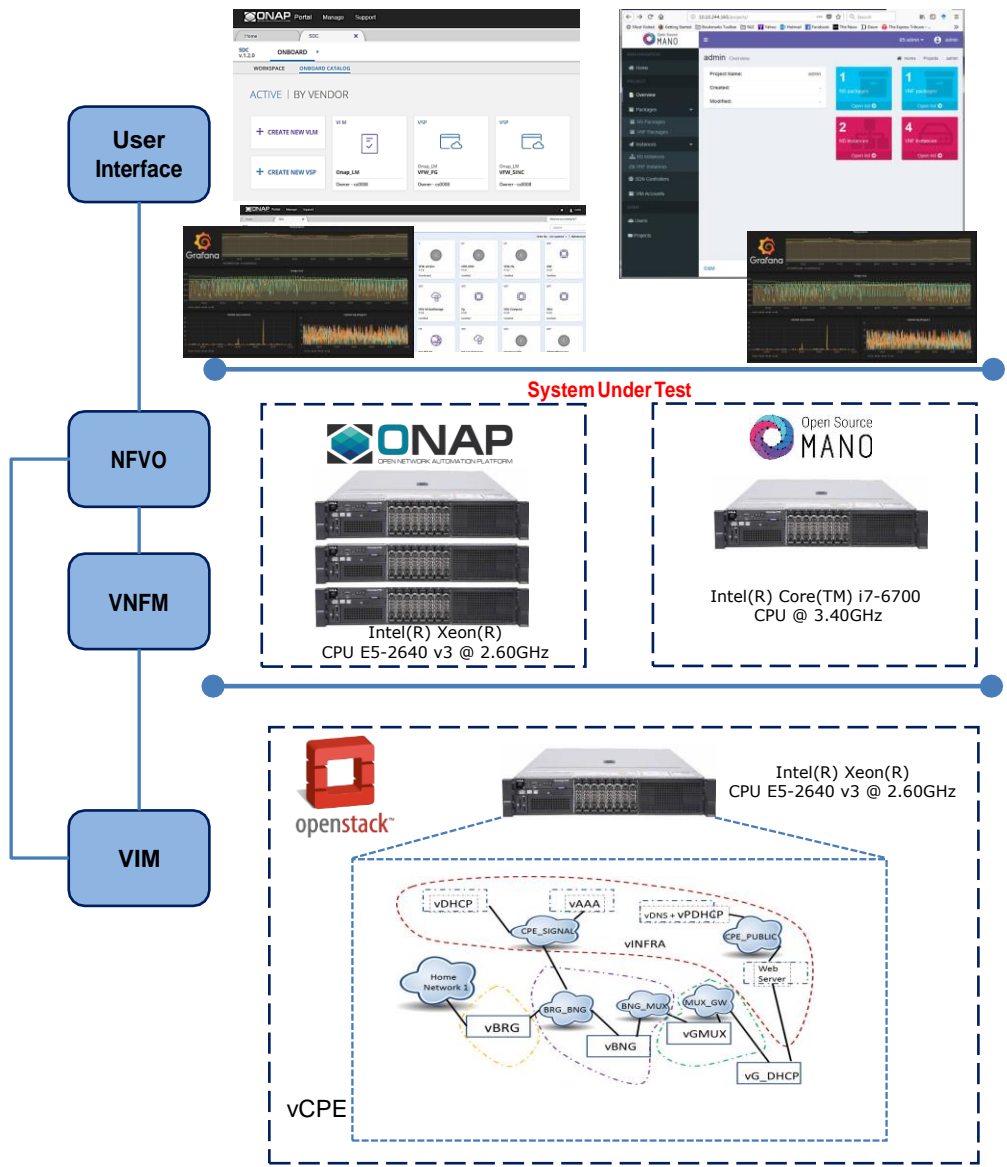

Figure 3: Overview of Our Testbed Setup ONAP-B and OSM-4 With Common OpenStack VIM.

### A. Benchmarking Testbed Setup

For our analysis we use the ONAP-Beijing release (ONAP-B) [3] and OSM Release 4 (OSM-4) [4]. In order to ensure common test environment on the basis of which to make fair and comparative analysis we use a common VIM framework and a common VNF type for both ONAP-B and OSM-4 testbeds. For the VIM we used OpenStack Ocata version [10] while for the VNF we used the virtual Customer Premises Equipment (vCPE) from ONAP [22], the details of which will be provided later. We also used rack-mounted bare-metal servers to replicate an operator's data-center environment. Fig. 3 gives an overview of the layout of the two isolated test-beds using a common VIM based on OpenStack, which is supported by both MANO platforms. Table I provides an overview of the VIM platforms supported by ONAP-B and OSM-4. Further details of the respective test-beds are given in the following sub-sections.

Table I: Supported VIM platforms by OSM-4 and ONAP-B

| Supported VIM Platforms | | | | |
|---|---|---|---|---|
| MANO Platform | Openstack | Vmware | AWS | OpenVIM |
| OSM-4 | ✓ | ✓ | ✓ | ✓ |
| ONAP-B | ✓ | X | X | X |

```bash
#!/bin/bash
wget https://osm-download↲
 ↪.etsi.org/ftp/osm-4.0-four/install_osm.sh
chmod +x install_osm.sh
./install_osm.sh --elk_stack --pm_stack --vimemu
export OSM_HOSTNAME=127.0.0.1
export OSM_SOL005=True
docker stack ps osm|grep -i running
docker service lsdocker stack ps osm|grep -i
 ↪running
docker service ls
osm vim-create --name openstack-VIM --user admin
 ↪ --password xxxx --auth_url
 ↪http://10.10.xx.xx:5000/v2.0 --tenant admin
 ↪ --account_type openstack
 ↪ --config='{security_groups: default, keypair:
 ↪secret-key-osm}'
#Deploy vCPE demo
#Call get_list_of_vnfs.sh and collect measurements
```

Listing 1: OSM Deployment & OpenStack VIM Attachment

**OSM-4 Test-bed Setup.** In the following we describe the steps for the OSM testbed deployment as summarized in Listing 1. For setting up the OSM-4 test-bed, we used a single Intel(R) Core(TM) i7-6700 CPU @ 3.40GHz server with Ubuntu 16.04 operating system. The base OSM-4 MANO system is composed of 10 functional modules installed in 10 Docker containers [4]. Additional 3 Docker containers are required when installing the optional Performance Management (PM) and Fault Management (FM) services, which we did. The installation were verified by checking the number of active docker containers running. The PM service also offers the Prometheus time-series database and the Grafana service for visualization.

The next task after having OSM-4 up and running is to attach an OpenStack VIM so as to deploy VNFs. For the VIM we use another rack mounted server based on Intel(R) Xeon(R) CPU E5-2640 v3 @ 2.60GHz installed with the 64-bit variant of Ubuntu 16.04 and running OpenStack Ocata release. Note that this server is also serving as an NFVI compute node for deploying the VNFs. We also setup Ceilometer, Gnocchi and Aodh alarm and notification services on the OpenStack VIM so as to profile and monitor the VNFs. The Aodh alarm service is intended to be used when testing LCM operations. Fig. 3 shows the reference architecture of our deployment.

**ONAP-B Testbed Setup.** In the following we describe the steps for the ONAP testbed deployment as summarized in Listing 2. The ONAP-B framework consists of 20 functional modules. The functionality of each module is implemented using multiple containerized services based on Docker [23]. For example, the Data Collection, Analytics, Events (DCAE) functional module is composed of 13 distributed containerized services [24]. There are two prescribed options for installing ONAP-B. One option is to use Kubernetes (K8S) [25] where the containerized functions are distributed within pods (i.e., group of containers). The second option is to install each of the 20 functional components in individual VMs, where each VM will contain a cluster of containers that makes up that specific functionality (e.g., DCAE, CLI, SDC etc).

During the installation process, the K8S-native installation option was not successful due to multiple container/component failure. The K8S orchestrator would go into indefinite loops in order to recreate the containers. This was a widely reported

```
#OpenStack Deployment
    Step1: Prepare OpenStack Cloud with
            minimum of 88vCPU, 178GB RAM, 1.76TB
            ↪Storage
    Step2: Make sure: Cinder, Glance, Heat, Horizon,
        ↪ Keystone, Neutron, Nova are installed in your
        ↪ OpenStack
end
#OpenStack Deployed
    Step3: Make sure to have at least 23 floating
       ↪IP addresses
    Step4: Connect your OpenStack to Internet
    Step5: Prepare a public SSH key to access the
       ↪various VMs
    Step6: Prepare Ubuntu 14.04  and 16.04 images
       ↪and create OpenStack Image
    Step7: Prepare Set of OpenStack flavors:small,
       ↪medium, large, xlarge
    Step8: Get Template file:onap_openstack.yaml
       ↪Environment file: onap_openstack.env
    Step9: Update onap_openstack.env with
            your OpenStack information
            public_net_id: NETWORK ID/NAME
            pub_key:       PUBLIC KEY HERE
            openstack_tenant_id:OPENSTACK PROJECT ID
            dcae_deployment_profile: R2
            ......
    Step10: source your OpenStack openrc.sh
    Step11: Create a stack with: openstack stack
      ↪  create -t onap_openstack.env -e
      ↪  onap_openstack.env  STACK_NAME(ONAP)
    end
# VMs Ready
    Step12: Verify Installation
        Step12.1 SSH to onap-robot
        Step12.2 Run the health check command
           ↪docker exec -it
           ↪openecompete_container/var/opt/
           ↪OpenECOMP_ETE/runTags.sh
    Step13: Register OpenStack VIM
        curl -X PUT \
          'https://<AAI_VM1_IP>:8443/aai/v11/cloud↲
            ↪-infrastructure/cloud-regions/cloud↲
            ↪-region/CloudOwner/RegionOne'
            ↪ \
          -H 'accept: application/json' \
          ............
          -d '{
            "cloud-type":"openstack",
            "owner-defined-type":"t1",
            "cloud-region-version":"ocata",
            "identity-url":"<keystone auth url,
            ........
            "tenants": {
                "tenant": [
                    {
                        "tenant-id":"xxxxxxxxxx",
                        "tenant-name":"vCPE"
                    }
                ]
            },.....}
    Step14: Deploy vCPE demo
    Step15: Call get_list_of_vnfs.sh and collect
       ↪ measurements
end
```

Listing 2: ONAP Deployment & OpenStack VIM Attachment

issue and thus this option was not possible in ONAP-B.

We thus adopted the second option, which was to install the ONAP-B components into individual VMs using the OpenStack HEAT Orchestration Template (HOT). For this purpose, it is first required to create a private OpenStack cloud comprising 1 controller and 2 compute nodes. The controller runs the OpenStack services such as Keystone, Heat, SWIFT and Nova compute services, whereas the compute nodes provide the resources and running the Nova compute service. For this we used 3 rack mounted servers based on Intel(R) Xeon(R) CPU E5-2640 v3 @ 2.60GHz installed with the 64-bit variant of Ubuntu 16.04 and running OpenStack Ocata release.

Table II: vCPE VNF components resource footprint.

| Use Case VNFs | VNF Compon. | Flavor | vCPUs | Memory (MB) | Storage (GB) |
|---|---|---|---|---|---|
| Residential Broadband vCPE | vBNG | m1.medium | 2 | 4096 | 40 |
| | vGDHCP | m1.medium | 2 | 4096 | 40 |
| | vBRG | m1.medium | 2 | 4096 | 40 |
| | vGMUX | m1.medium | 2 | 4096 | 40 |
| | vInfra | m1.medium | 2 | 4096 | 40 |

The HEAT script instantiates 20 VMs, configured each VM with a respective ONAP-B functional component by clustering the necessary containerized functions to make up that particular functional component. The HEAT script also enabled the necessary networking between the functional components. The sanity of the installation was verified by the ONAP-B "Robot" functional component, which is part of the 20 functional components making up ONAP-B. For monitoring and data acquisition we used the in-built DCAE functions, such as Prometheus time-series database and Grafana for visualization.

After the successful installation of ONAP-B, an external VIM is attached to the ONAP-B orchestrator using the VIM-Manager functional component of ONAP-B . The specification and properties of the VIM is the same as that for OSM-4 described above.

**Virtual CPE (vCPE) VNFs.** For analyzing and comparing the supported features and performance of ONAP-B and OSM-4 we used vCPE as a common use case. The vCPE is a complex VNF composed of 5 VNF components (VNFC) and provided by ONAP [22]. For both test-beds the vCPE VNFs are installed on the compute node managed by the respective VIM. Each vCPE VNFC is on-boarded and instantiated on a separate VM and each has a specific resource footprint. Table II provides the summary of the vCPE VNFCs and their respective footprint. Since the vCPE VNF is provided by ONAP-B as an example VNF, it came pre-packaged with the required images, NSD and VNFD. However, in order to deploy the vCPE use-case in OSM-4 environment we had to re-write the VNFD and NSD in order to comply with the OSM VNF packaging guidelines. That is, to package the vCPE software image in QCOW2 format. Also the VNFD and NSD had to be modified to comply with the OSM-4 test-bed networking environment. Afterwards, the vCPE software image is uploaded to the OpenStack Glance image service, whereas the VNFD and NSD were uploaded to the OSM-4 system repository. This also verified the ability of OSM-4 to on-board third party VNFs.

**OSM & ONAP VNFs Metrics Collection.** After the vCPE VNFs are deployed in the OpenStack VIM, for metrics collection we used Gnocchi [26] project which is a project under the Ceilometer [27] program. The Ceilometer project stands deprecated because of performance issues due to large amount of data being collected and at some point without having the data storage back-end collapsing. Gnocchi solves this problem by using ceilometer as umbrella and introducing a new way to store and aggregate time series data and provides APIs (multi-tenant time-series metrics as a service) to collect important metrics about a VNF deployed in OpenStack. Listing 3 shows our python script to collect metrics for VNFs deployed and running in OpenStack using Gnocchi services.

```python
import json
import os
from pandas import Series
def main():
    os.system('source demo-openrc.sh')#Authenticate
     ↪with OpenStack
    #Call get list of vnfs and save name and id into
     ↪vnfs_list.json
    get_vnf_list='openstack server list -c ID -c
     ↪Name'
    +'-f json >  vnfs_list.json'
    os.system(get_vnf_list);
    #read list of VNFs from vnfs_list.json
    vnf_list=json.load(open('vnfs_list.json'));
    #Iterate over the list of VNFs
    for vnf in vnf_list:
        vnf_id=vnf.get('ID');   #read VNF ID
        vnf_name=vnf.get('Name') ;  #read VNF Name
        #get metrics resources available for the VNF
         ↪using ID and Save the response to
         ↪vnf_name.json
        vnf_resources='gnocchi resource show '+
         ↪vnf_id+' '+'-f json > '+' '+
         ↪vnf_name+'.json';
        os.system(vnf_resources);#execute command on
         ↪terminal
        vnf_resource_json=json.load(open(vnf_name
         ↪+'.json'));  #read saved vnf_name.json
        vnf_resource_metrics=
         ↪vnf_resource_json.get('metrics');#read
         ↪metrics available for the VNF
        #Iterate over each metrics
        for metrics in vnf_resource_metrics:
            unit=vnf_resource_metrics.get(metrics)
             ↪#read unit of metrics
            #read usage statistics of the VNF for
             ↪the current metrics
            usage='gnocchi measures show '+unit
             ↪+' '+'-c timestamp --utc  -c
             ↪value '+' '+' -f csv$'
            os.system(usage);#execute command on
             ↪terminal
            #save measured value into csv file
             ↪vnf_name_metrics.csv
            series=Series.from_csv(vnf_name+'_'
             ↪ +metrics+'.csv', header=0)
if __name__ ==' main ':
     main()
```

Listing 3: OSM & ONAP VNFs Metrics Collection

*B. Performance Analysis*

In the following sub-sections, we present the performance analysis and comparison of ONAP-B and OSM-4 based on the functional and operational KPIs, and also share our observations during the process.

**Functional KPIs-based analysis.** Table III provides a comparison of the resource footprint for installing OSM-4 and ONAP-B. As can be seen, OSM-4 platform requires significantly less resources than ONAP-B. For example OSM needs about 2.27% of vCPU, 4.5% of RAM, 2.27% of storage and 4.43% of IP addresses of that of ONAP, hence with such a small resource footprint, OSM is suitable for deployment in a computing environment with limited resources, such as Edge Data-centers.

Also in terms of the variety of VIMs that a MANO platform can support, it is seen from Table I that OSM-4 claims to support 4 different variety of VIM frameworks, whereas ONAP-B supports only 1 i.e., OpenStack. It should be noted that we did not have the resources to test OSM-4 claim for supporting other VIM platforms except OpenStack. Also owing to the limited size of our test-bed we could not verify the max number of VIMs and VNFs that the respective MANO platform could effectively manage. Such verification can only be made in medium to large scale Datacenters with a large resource pool. Also, in our experience we found that both OSM-4 and ONAP-B do not support containerized VNFs, and there is no information as to when such a support will be available. Containerized version of VNFs are important for edge Datacenters with limited compute resources.

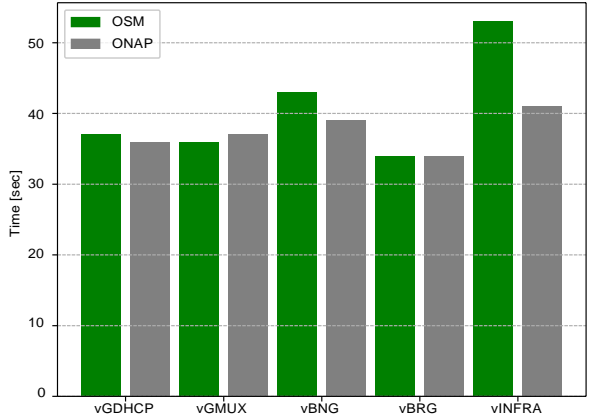

(a) OPD for vCPE VNF Components. VNFs.

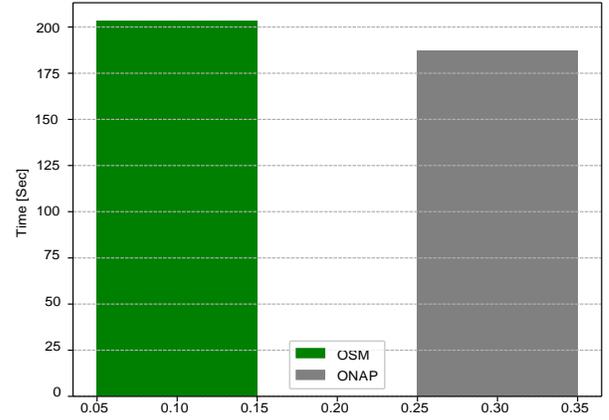

(b) Aggregate OPD. Time.

Figure 4: OSM-4 vs ONAP-B On-boarding Processing Delay (OPD) for vCPE VNFs.

Table III: OSM vs ONAP resource footprint comparison.

| Resource | OSM-4 | ONAP-B |
|---|---|---|
| vCPU | 2 | 88 |
| Memory(GB) | 8 | 176 |
| Storage(GB) | 40 | 1760 |
| IP Addresses | 1 static | 20 Floating 3 static |

**Operational KPIs-based Analysis.** Here we present our performance comparison analysis of both platforms under the common vCPE use-case based on the KPIs defined in Section III-B, i.e, On-boarding Process Delay(OPD) and Deployment Process Delay (DPD). Run-time Orchestration Delay (ROD) and Quality of Decision (QOD) could not be evaluated since run-time orchestration operations were not yet supported in both platforms. We also measure the CPU and memory utilization for the vCPE VNFCs in both the test-beds.

Fig. 4a measures and compares the OPD of the individual vCPE VNFCs for OSM-4 and ONAP-B, while Fig. 4b compares the aggregate OPD for the two platforms. Although at an aggregate level, the overall difference is not significant (see Fig. 4b, but at the VNFC level (see Fig. 4a) there are some for which the OPD is higher in OSM-4 than in ONAP-B and vice versa. It is observed that the OPD for VNFCs with complex services and network requirements such as vINFRA and vBNG have generally a high OPD as compared to the other vCPE VNFCs. This is due to the time it takes to prepare and boot a VM for such VNFCs. It may be noted that OSM in general incurs higher OPD than ONAP. Although this difference in the OPD incurred by OSM and ONAP may not be significant for different VNFCs, but then this difference can get compounded in case of a large scale deployment consisting of 100s of 1000s of VNFCs.

In Fig. 5a, we see a similar behavior when comparing and analysing the DPD incurred by OSM and ONAP when instantiating the respective vCPE VNFCs. Although at the aggregate level (see Fig. 5b) the DPD incurred by both OSM-4 and ONAP-B is similar, but then at the individual VNFC level it may vary; depending again on the service/resource complexity of the VNFC that is being instantiated on a deployed VM. Again for some VNFC the OSM-4 may incur a higher DPD, while for others it is ONAP-B. As mentioned above, although the individual difference may not be dramatically significant but then it gets compounded for dense deployment scenario.

Regarding CPU utilization, the OSM-4 records a higher utilization than ONAP-B as seen in Fig. 6b, but at the individual VNFC level it may vary depending on the VNFC. For example as can be seen from Fig. 6a OSM-4 incurs a much higher CPU utilization for vGMUX and vBNG, whereas for vGDHCP ONAP-B incurs a higher CPU utilization.

Since we did not have the means to capture the process-protocol messages exchanged between the OSM-4, ONAP-B and VIM, we can only infer that the difference in the OPD and DPD for the OSM-4 ad ONAP-B platforms is due to the difference in the internal processes of the respective platforms. This inference is supported by Fig. 7a and Fig. 7b where the memory utilization, both at the individual VNFC level and the aggregate level, is almost the same for OSM-4 and ONAP-B. This is because both are attached to the same OpenStack managed compute infrastructure.

Regarding run-time LCM operations, both OSM-4 and ONAP-B claimed that they support Scaling operation only.

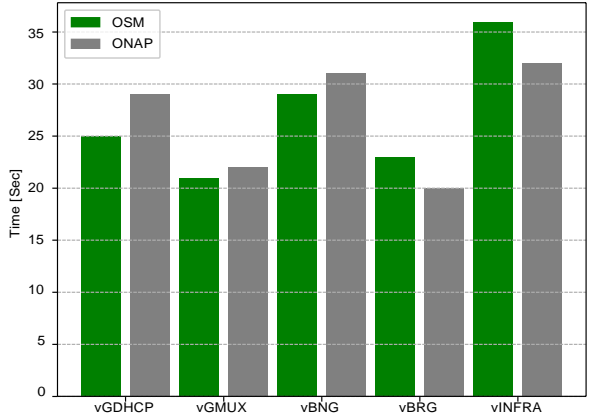
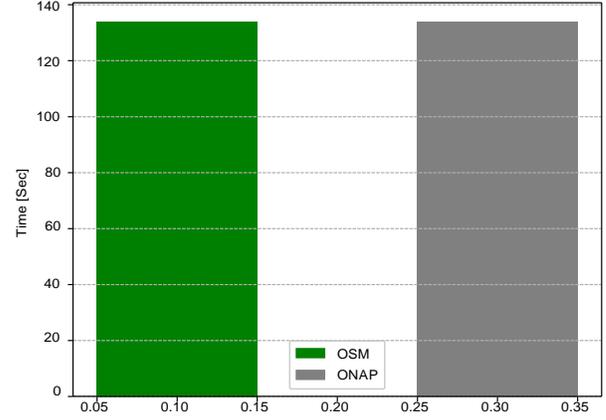

(a) DPD for vCPE VNFCs.

(b) Aggregate DPD Time.

Figure 5: OSM-4 vs ONAP-B VNF Deployment Process Delay (DPD) Comparison.

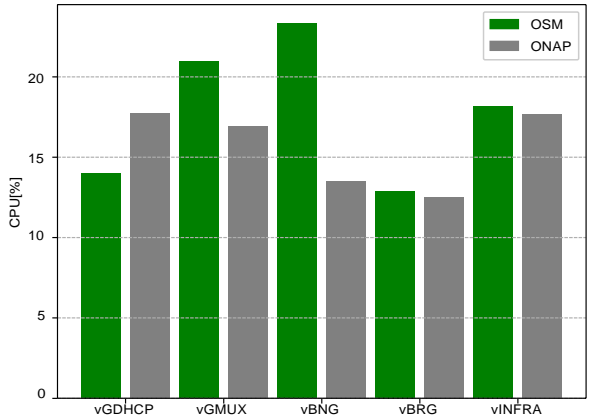
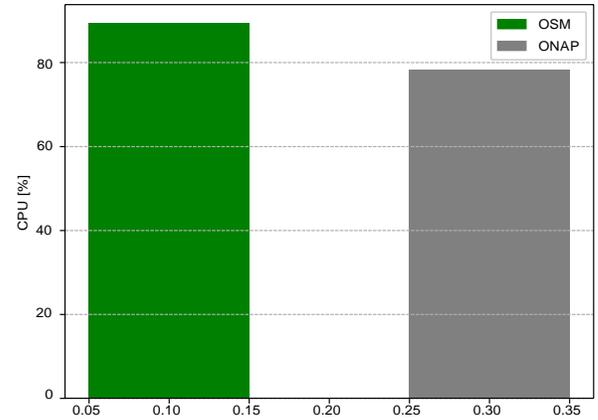

(a) CPU Usage of vCPE VNFs.

(b) Aggregate CPU Usage.

Figure 6: OSM vs ONAP VNF CPU Usage Comparison.

However, we discovered that such LCM operations are not supported by the respective platforms despite the claims. It is for this reason we were unable to analyze the run-time operation performance based on ROD and QoD KPIs.

In summary when it comes to DPD of VNF components our experimental results shows that, although at an aggregate level there is no significant difference (see Fig. 5b), but then there are differences at the individual VNFC level. For example, as shown in Fig. 5a individual VNFCs such as vINFRA, which comprises of multiple services have a longer DPD time than other less complex VNFCs. Fig. 6a also shows the CPU usage for individual VNFCs, here it can be clearly seen that there are variations on individual VNFCs, interestingly on aggregate level Fig. 6b shows that OSM-4 deployed VNFCs consume more CPU resources. Nonetheless, memory usage both on individual and aggregate level seems uniform on both platforms as depicted in Fig. 7a and Fig. 7b because both test-beds use a common VIM platform.

We have also provided a comparison matrix of OSM-4 and ONAP-B in Table IV summarizing different aspects of the two MANO platforms that also reflects our experience/observations. For instance, in our experience we found OSM-4 to have a comparatively easier learning curve than that for ONAP-B. By learning curve we mean that the time and effort required to setup the platform, attach a VIM etc. In terms of Multi-user support, ONAP-B provides a well structured support for different user roles as compared to OSM-4. ONAP- B also offers multiple installation methods including cloud native installation support with Kubernetes (although error- prone), while OSM-4 has a much better CLI client that is designed following the OpenStack CLI client pattern. Thus anyone who has experience with the OpenStack CLI will find using the OSM CLI client intuitive and convenient. In addition to that OSM-4 has also low resource footprint and multiple VIM support compared to ONAP-B as discussed before. OSM-4 also provides the flexibility to link with 3rd party monitoring system whereas ONAP-B has a built-in sophisticated DCAE engine for the purpose of monitoring and analysis. Moreover, both platforms, despite the claims, don't have functioning support for basic LCM operations such as scaling, thereby making OSM and ONAP not yet mature enough for deployment in production grade environment.

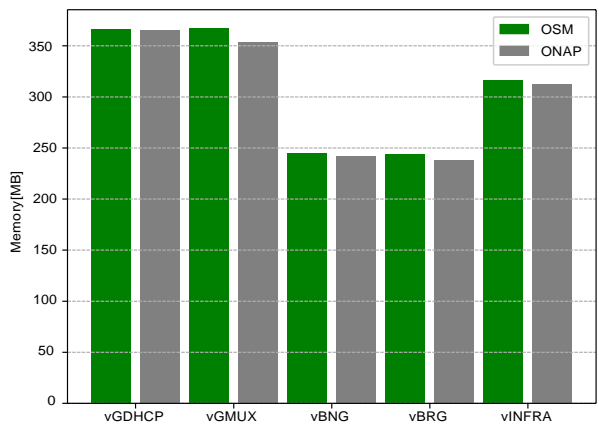
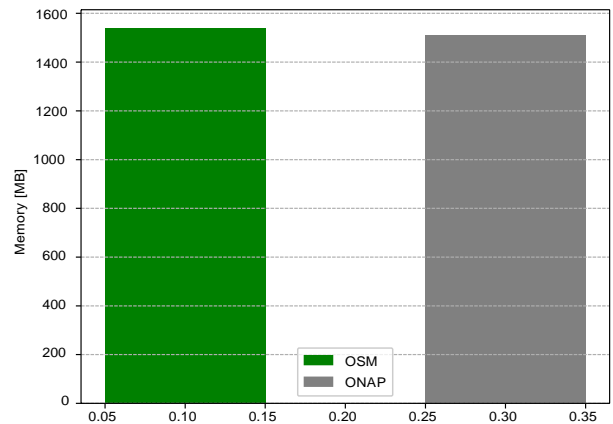

(a) Memory Usage of vCPE VNFs.  (b) Aggregate Memory Usage of vCPE VNFs.

Figure 7: OSM vs ONAP VNF Memory Usage Comparison.

Table IV: OSM-4 vs ONAP-B Comparison Matrix

| Evaluation Criteria | OSM-4 | Remarks | ONAP-B | Remarks |
|---|---|---|---|---|
| Resource footprint | Low | See Table III | High | Needs high spec rack-mounted servers (See Table III) |
| Bare metal server Installation support | ! | Easier for research & development | X | Not possible to run ONAP in a single PC |
| Kubernetes Installation support | X | Not supported yet | ! | ONAP-B can have multiple instances with different name-spaces |
| Performance Monitoring | ! | Open for 3rd party monitoring services | ! | DCAE module of ONAP is responsible for this with much richer APIs for developers of Data Analytics Applications |
| Multi-VIM Support | ! | OpenStack, VMware, AWS | X | Only OpenStack |
| CLI Support | ! | Powerful CLI, seems to follow OpenStack CLI paradigm | ! | Not user friendly |
| LCM Support | X | Not available | X | Not available |
| Learning Curve | Easy | Very good wiki & active community support through Slack | Hard | Not well documented, pretty difficult to get clear information. |
| Multi-User Support | ! | Can be improved | ! | With different roles for Designer, tester, governor and operator |
| Multi-Site Support | ! | Not tested | ! | Not tested |

## V. CONCLUSIONS

In this paper, we have motivated the need to analyze the performance of NFV MANO systems and highlighted the challenges of such a task. For this purpose, we have specified a set of Functional and Operational KPIs that can be used to benchmark the performance of an NFV MANO system. Based on the defined KPIs, we analysed the performance of two popular and competing open-source NFV MANO platforms, namely OSM-4 and ONAP-B. To ensure a fair comparison, a common VIM platform based on OpenStack was used over which the same type of of VNF were utilized i.e., a vCPE VNF. Based on the described setup, we have provided valuable insights regarding both platforms, including wide feature-gaps between what was claimed and what actually worked in practice. According to our evaluation, both platforms can successfully on-board and instantiate VNF and NS instances. However, run-time orchestration actions such as scaling operations failed on both platforms. Thus the performance of the two platforms in terms of ROD and QoD KPIs could not be analysed and compared. The performance of the two platforms have been compared in operational terms (resources footprint, on-boarding process delay - OPD and deployment process delay - DPD) as well as in functional terms (variety of supported VIM platforms, scaling support, maximum number of VNFs). In Table IV, we have also presented a comparison matrix of OSM-4 and ONAP-B from different perspectives.

## VI. FUTURE WORK

At the time of compiling this paper new versions of OSM and ONAP got released, namely OSM Release 5 (OSM-5)

and ONAP Casablanca Release (ONAP-C). New features have been added while other non-functioning features of OSM-4 and ONAP-B are claimed to have been rectified, such as the *scaling* operation.

The claimed resource footprint of OSM-5 and ONAP-C, is shown in Table V. As it can be observed, ONAP-C still has a significantly large resource footprint as compared to OSM-5. Even with the container-based option (i.e., Kubernetes) ONAP-C resources footprint continues to dominate. On the other hand, OSM-5 claims that it has brought down its Memory and Storage requirements by 50%. In addition, ONAP-C claims it has expanded its support from OpenStack to other VIMs like VMware and Wind River's Titanium.

Table V: OSM-5 vs ONAP-C resource footprint

| Orchestrator | vCPU | Memory (GB) | Storage (GB) | Installation Mode |
|---|---|---|---|---|
| OSM-5 | 2 | 4 | 20 | Bare-metal server(VM) |
| ONAP-C | 88 | 176 | 1760 | OpenStack |
|  | 112 | 224 | 160 | Kubernetes |

In our future work we plan to upgrade our respective test-beds to these latest releases and analyse, in addition to the OPD and DPD KPIs, the run-time orchestration decision (ROD) and quality-of-decision (QOD) KPIs. This will allow us to perform both stress and reliability tests on the respective MANO platforms.

Table VI: Abbreviations List

| Abbreviation | Meaning |
|---|---|
| **BSS** | Business Support System |
| **DCAE** | Data Collection and Analytics Engine |
| **DPD** | Deployment Process Delay |
| **EM** | Element Manager |
| **FCAPS** | Fault, Configuration, Accounting, Performance, Security |
| **FM** | Fault Management |
| **HOT** | Heat Orchestration Template |
| **KPIs** | Key Performance Indicators |
| **LCM** | Life Cycle Management |
| **MANO** | Management and Orchestration |
| **NFV** | Network Function Virtualization |
| **NFVI** | Network Function Virtualization Infrastructure |
| **NFVO** | Network Function Virtualization Orchestration |
| **NMS** | Network Management System |
| **NS** | Network Service |
| **NSD** | Network Service Descriptor |
| **ONAP** | Open Network Automation Platform |
| **OPD** | On-boarding Process Delay |
| **OSG** | Open Source Group |
| **OSM** | Open Source MANO |
| **OSS** | Operation Support System |
| **PM** | Performance Management |
| **QoD** | Quality Of Decision |
| **QoS** | Quality Of Service |
| **ROD** | Run-time Orchestration Delay |
| **SDC** | Service Design Center |
| **SLA** | Service Level Agreement |
| **TRL** | Technology Readiness Level |
| **vBNG** | Virtualized Bridge Network Gateway |
| **vBRG** | Virtualized Bridge Residential Gateway |
| **vCPE** | Virtual Customer Premises Equipment |
| **vCPU** | Virtual CPU |
| **vEPC** | Virtual Evolved Packet Core |
| **vGDHCP** | Virtualized Gateway and DHCP |
| **vGMUX** | Virtualized Gateway Multiplexer |
| **VIM** | Virtual Infrastructure Manager |
| **vINFRA** | Virtual Infrastructure |
| **VL** | Virtual Link |
| **vLB** | Virtual Load Balancer |
| **VMs** | Virtual Machines |
| **VNFD** | Virtual Network Function Descriptor |
| **VNFFG** | VNF Forwarding Graph |
| **VNFM** | Virtual Network Function Manager |
| **VNFs** | Virtual Network Functions |